\documentclass{egpubl}
\usepackage{eurovis2021}

\usepackage[T1]{fontenc}
\usepackage{dfadobe}  

\usepackage{cite}  
\BibtexOrBiblatex
\electronicVersion
\PrintedOrElectronic
\ifpdf \usepackage[pdftex]{graphicx} \pdfcompresslevel=9
\else \usepackage[dvips]{graphicx} \fi

\usepackage{egweblnk}

\title[Enhancing Reading Strategies]%
      {Enhancing Reading Strategies by Exploring \linebreak A Theme-based Approach to Literature Surveys}

\author[T. Howden, P. Le Bras, T. S. Methven, S. Padilla \& M. J. Chantler]
{\parbox{\textwidth}{\centering T. Howden$^1$,
        P. Le Bras$^{1}$\orcid{0000-0002-0670-6552},
        T. S. Methven$^{2}$\orcid{0000-0002-9518-3596},
        S. Padilla$^{1}$\orcid{0000-0002-7104-8349},
        and M. J. Chantler$^{1}$\orcid{0000-0002-8381-1751} 
        }
        \\
{\parbox{\textwidth}{\centering $^1$School of Mathematical and Computer Sciences, Heriot-Watt University, Edinburgh, United Kingdom\\
         $^2$School of Computing, Edinburgh Napier University, Edinburgh, United Kingdom
       }
}
}

\begin{document}


\maketitle
\begin{abstract}
   Searching large digital repositories can be extremely frustrating, as common list-based formats encourage users to adopt a convenience-sampling approach that favours chance discovery and random search, over meaningful exploration. We have designed a methodology that allows users to visually and thematically explore corpora, while developing personalised holistic reading strategies. We describe the results of a three-phase qualitative study, in which experienced researchers used our interactive visualisation approach to analyse a set of publications and select relevant themes and papers. Using in-depth semi-structured interviews and stimulated recall, we found that users: (i) selected papers that they otherwise would not have read, (ii) developed a more coherent reading strategy, and (iii) understood the thematic structure and relationships between papers more effectively. Finally, we make six design recommendations to enhance current digital repositories that we have shown encourage users to adopt a more holistic and thematic research approach.

\ccsdesc[300]{Human-centered~Human computer interaction (HCI)}
\ccsdesc[300]{Human-centered~Visualization}
\ccsdesc[300]{Applied computing~Collaborative learning}

\printccsdesc   
\end{abstract}  
\section{Introduction}

Large digital repositories of research papers and associated materials are ubiquitous and used on almost a day-to-day basis by many researchers \cite{berry1996digital} \cite{zha2015comparing}. These repositories combine accessibility of information and technology to enable users to instantly and conveniently search and access resources from diverse collections as described by Cherukodan \cite{cherukodan2013using}. As a result, these digital repositories are commonly used by researchers in their standard approach towards literature discovery and to facilitate their reading strategies; however, they present challenges and issues.

These repositories frequently use a keyword search to highlight resources that may be of relevance to the user; this method has been widely observed in current interfaces and broadly reported in research \cite{Acm} \cite{GoogleScholar} \cite{Watson} \cite{Springer}. A disadvantage of search methods is their reliance on the users’ expertise and previous knowledge of an area, this causes difficulties when users explore new domains as described by Kotchoubey et al. \cite{Kotchoubey2011P912CA} and Wilson et al. \cite{wilson2010keyword}, for example  when they don’t know what to search for, or in the case of concept homonymy (e.g. “neural network” in biology or computer science). Moreover, in these repositories, specific fields of information are quite prominent in the search result (e.g., title and author information) \cite{Acm} \cite{Sciencedirect} \cite{GoogleScholar} \cite{Microsoftacademic}; it is, however, unlikely that, for example, a title can adequately represent the whole content of the source. These disadvantages in current repositories and search methods increase the chance of users exploring irrelevant sources, advocating for a more time-consuming and frustrating trial and error approach, and being stuck at the start of their literature surveys, a situation commonly experienced by researchers.

To overcome these issues and challenges, we suggest using a top-down approach as inspired by Wilson et al. \cite{wilson2010keyword}, Padilla et al. \cite{Padilla2014hot} \cite{padilla2014british} \cite{padilla2017understanding} and in Le Bras et al. \cite{le2020visualising} work, where users begin by browsing an overview from a repository. Furthermore, Blei suggests that rather than finding new documents using traditional keyword search approaches, it would be better for users to take a theme-based approach to explore and digest collections \cite{blei2012probabilistic} \cite{blei2003latent}. We believe this behavior is a more natural solution to finding resources as it is common for literature sources to be created from a set of themes organized into a narrative. 

\begin{figure*}[tbp]
  \centering
  \mbox{} \hfill
  \includegraphics[width=.99\linewidth]{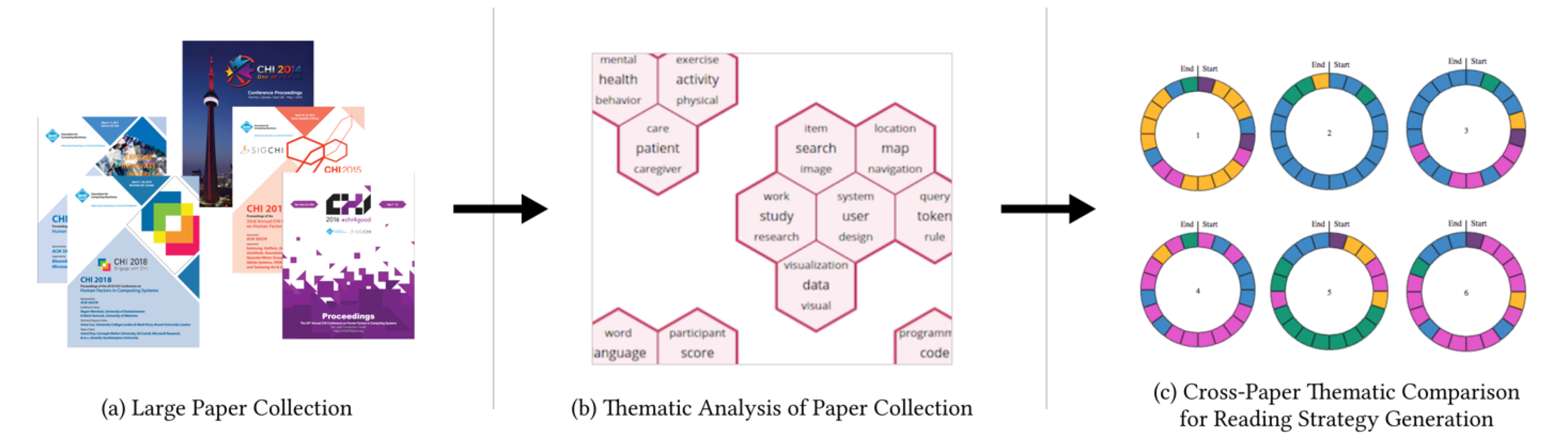}
  \hfill \mbox{}
  \caption{\label{fig:f1}%
           Summary of the proposed thematic methodology. We first run an analysis of a large collection of papers (a → b) to generate a theme-based overview. Participants can then explore themes and their relevant papers. Moreover, participants can investigate each of the papers using visual wheels, highlighting sequences and amounts for different themes. Finally, participants can compare selected resources to enable them generate an enhanced cross paper reading strategy (b → c).   }
\end{figure*}

In this paper, we explore user behaviors using thematic analysis tools along with data visualization techniques to see if we can visualize firstly, theme-based overviews of a paper collection to enable objective browsing and paper selection, and secondly, if visualizing sequences and quantities of themes within individual papers in a paper set aids the generation of a holistic cross-paper reading strategy. We conduct our investigation using a three-phase qualitative study, a set of tools, and a new six-step thematic methodology, as summarized in Figure \ref{fig:f1} inspired by Shneiderman’s Visual Information Seeking Mantra \cite{Shneiderman1996eyes} and Wilson’s et al. \cite{wilson2010keyword} exploration of information work.
 
It should be emphasized that our objective is to explore whether taking a thematic approach to browsing and selecting research papers allows users to adopt a holistic approach to these tasks followed by developing a reading strategy. We are not exploring issues with usability and performance of the proposed thematic methodology compared against commonly-used searching techniques in digital repositories as we want to focus on the user behaviors, gathering insights, and suggesting possible add-on enhancements to current methods and tools.

\noindent
The contributions of this paper, in summary, are:
\begin{enumerate}
  \item We explore visual thematic tools and an associated methodology for the selection of a paper set and generation of a cross-paper reading strategy.
  \item We report insights on the effect of promoting thematic content, contrasted with the recalled experience of commonly-used title-based approaches and tools.
  \item We propose, from our results, six design recommendations (R1-R6) for enabling effective browsing and selection capabilities to improve users’ experience and enhance current tools.
\end{enumerate}

\section{Background and Related Work}

In this section, we look at current approaches for browsing and the selection of content from digital research repositories; we then discuss how visualizations can aid those tasks and motivate our proposed methodology.

\subsection{Browsing and selecting in digital repositories}

There are many different definitions of what is considered a digital repository, otherwise known as digital libraries. Chowdhury and Chowdhury \cite{chowdhury2003introduction} place digital repositories into two major categories based on Borgman’s discussions \cite{borgman1999digital}. These categories firstly look after collecting and organizing literature and secondly focus on accessing and retrieving these digital sources. In this paper, we concentrate on the latter and consider a digital repository to be an online platform that allows users to search and retrieve digital copies of literature sources.

These collections of resources are widely available from the publishers themselves \cite{Acm} \cite{Sciencedirect} \cite{Ieeexplore} \cite{Springer}. Additionally, companies such as Google and Microsoft provide search engines reaching multiple repositories \cite{GoogleScholar} \cite{Microsoftacademic}. All of these platforms integrate the same core mechanism for browsing, that is, using keywords as the basis of the search, with the ability to then filter results using facets such as date published, authors, institutions and publication type \cite{Sciencedirect} \cite{xie2016discover}. We believe that Shneiderman's Visual Information Seeking Mantra \cite{Shneiderman1996eyes} proposes another browsing mechanism: first offering an overview of a research area, then allowing the user to focus on particular themes, and finally giving access to the sources. A related approach has been partially implemented (Research Perspectives \cite{Researchperspectives}), its use, however, remains minor in comparison to the keyword search method. As a result, we believe more research is needed to explore the user behaviors to facilitate the use of such complementary approaches to common search mechanism.

Additionally, result listings majorly emphasize title and author information, leaving out the explanation for relevance, and in turn the order in which results appear. Beel and Gipp found from reverse engineering techniques that the ranking algorithm by Google Scholar \cite{GoogleScholar} used the number of citations as the highest weighted factor \cite{beel09}. They also found that the occurrence of search terms in the title outweighed their occurrence in the full text, making no difference to the ranking if the search term appeared only once or multiple times, thus presenting a biased representation of the source content \cite{beel2009google}. It also emphasizes difficulties in assessing the relevance of a source, given the prominence of attractive titles \cite{haggan2004research} \cite{russell2016importance}.   

Modern digital repository platforms have tried to visualize the theme of the papers using word clouds and similar abstractions \cite{Scopussearch} \cite{Watson}; however, the main emphasis of their mechanism still relies upon the search of title keywords to find resources. To our knowledge, there is a lack of research and tools that offers users the ability to see thematic overviews, to explore how much of their search term appears in sources, and that gauges the relevance of these to their interests. 

Finally, there is some work in the manual annotation of themes, for example using crowdsourcing techniques, ConceptScape allows the annotation of lecture videos to highlight the content of each section, resulting in the facilitation of content discovery \cite{liu2018conceptscape}. Similar results could be achieved with textual content, for example, using analytic hierarchy processes \cite{golden1989analytic} \cite{kato2014using}, or systematic literature reviews \cite{nightingale2009guide} \cite{xiao2019guidance}. These methods are; however, time-consuming. Topic modeling\cite{blei2012probabilistic}, and in particular Latent Dirichlet Allocation (LDA) \cite{blei2003latent}, offers a time-efficient and effective method for uncovering and annotating the thematic structures within documents. Such methods have successfully been applied by Zhao et al. in the context of MOOC video repositories \cite{zhao2018flexible}.

\subsection{Facilitating browsing and selection with visualizations}

Popular digital repositories, such as Google Scholar \cite{GoogleScholar} or ACM DL \cite{Acm}, are heavily text-based, with limited amounts of imagery. Cognitive style research \cite{blazhenkova2009new} \cite{richardson1977verbalizer} \cite{thomas2010cognitive} suggests that visual users may not be using these text-based environments to their full potential. Therefore, being able to visualize literature sources, with a focus on themes and thematic structures, could better cater to these users preferred style of information presentation. Morris et al. \cite{morris2018understanding} demonstrated this with dyslexic users, where the interviewees reported a preference for interface uncluttered from substantial textual content. Besides, data visualizations and pictorial representations allow for better recall \cite{card1999readings} \cite{Nelson}; this highlight why techniques like icons and logos are used rather than text \cite{norman1995psychopathology}.

Notable work has been done to visualize search results rather than using text-based lists. WebSearchViz incorporates a circular interface to show result relevance in terms of how close they are to the center point of the circle \cite{10.1109/TVCG.2006.111}. TileBars shows the length of each result, highlighting the frequency of the search term \cite{hearst1995tilebars}. PubCloud presents a word cloud to summarize each of the listed results \cite{kuo2007tag}. Others, like LineUp, explicitly highlight how a result relates to each facet to explain a ranked list \cite{gratzl2013lineup}. Each of these designs provides suitable solutions to the problem of unexplained ranked lists of titles. Systems such as PivotPaths and PivotSlice aim to allow exploration of information resources to reduce the requirement for user-defined keywords \cite{dork2012pivotpaths} \cite{zhao2013interactive}. MOOCex presents a collection of educational videos using a Voronoi diagram to highlight the similarity between different videos in order to recommend a sequence of coherent videos to watch \cite{zhao2018flexible}.

Work has been done incorporating visual representations of topics to enable users to analyze \cite{oelke2014comparative} \cite{dunne2012rapid} \cite{padilla2013intuitive} and compare \cite{diakopoulos2015compare} \cite{alexander2015task} documents in a collections; however, to our knowledge, nobody has focused on using a theme-based approach to give an overview of a large collection of resources, or using this same approach for analyzing and comparing sources to generate a reading strategy.

We believe that visual representations of collections and individual sources with a thematic emphasis could allow the users to reflect and recall back to these representations, assisting with their browsing and selection tasks. Additionally, as we will be visualizing sequences of themes to describe the progression of content in a research paper, we have found work has been done on visualization for sequences. MatrixWave \cite{zhao2015matrixwave} visualizes the sequence of clickstream events on a website. Sankey diagrams are also commonly used to visualize sequences of objects \cite{riehmann2005interactive}. We found that although these are novel ways of presenting sequences, we wanted a representation that would allow for no training and intuitive interaction to allow users to find papers with common quantities of their selected themes.

\subsection{Summary}

There has been substantial work done on providing insights into ranked search results using data visualization techniques, including how similar each resulting item is to one another. To our knowledge, however, none of the existing solutions have entirely focused on using a visual theme-based approach to obtain a interactive visual overview of a large collection of resources, that can be filtered to facilitate comparison and analysis of a paper set, and that assist primarily with the generation of a holistic reading strategy. 

\begin{figure*}[tbp]
  \centering
  \mbox{} \hfill
  \includegraphics[width=.99\linewidth]{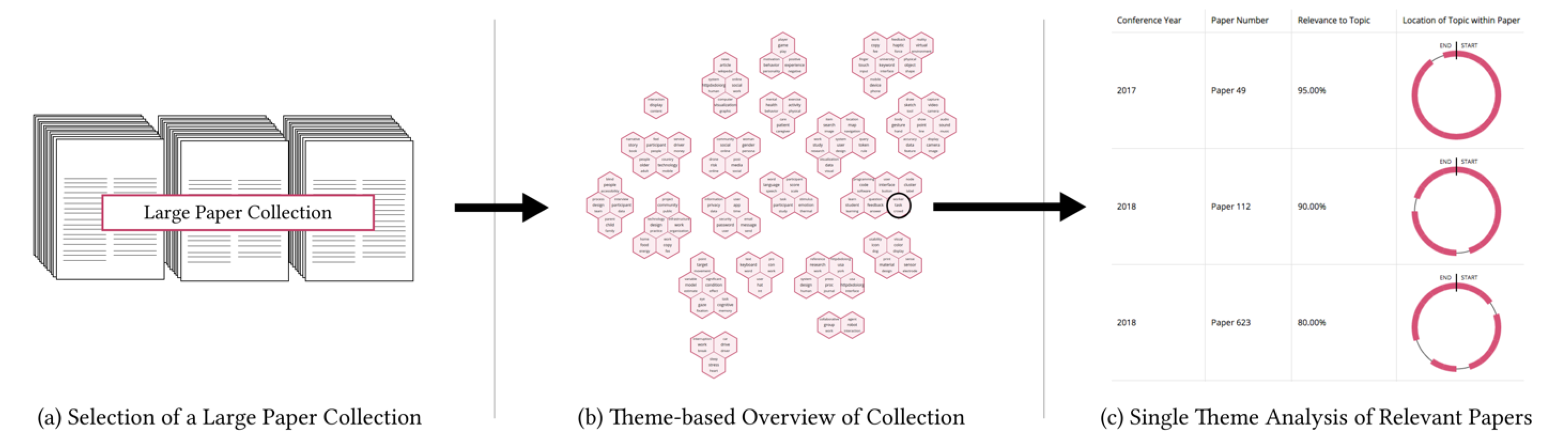}
  \hfill \mbox{}
  \caption{\label{fig:f2}%
           The first 3 stages of our thematic methodology. Tool 1 consist of (a) the selection of a large paper collection of information resources, (b) thematic analysis of the paper collection resulting in a theme-based overview of the content, and (c) single theme analysis highlighting the relevant papers based on where that theme appears in the source.}
\end{figure*}

\begin{figure*}[tbp]
  \centering
  \mbox{} \hfill
  \includegraphics[width=.99\linewidth]{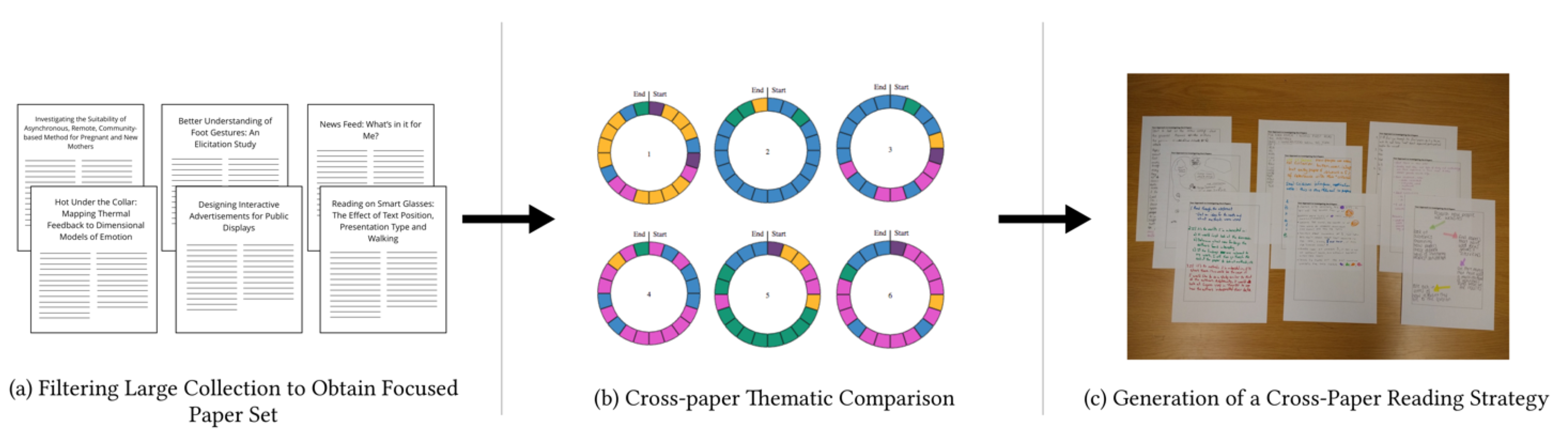}
  \hfill \mbox{}
  \caption{\label{fig:f3}%
           The final 3 stages of our thematic methodology. Tool 2 consists of (a) the focused paper set chosen by the user, (b) cross-paper thematic analysis of the paper set using theme wheels to represent sequences of themes from start to finish in each paper and, (c) generation of a cross-paper reading strategy.}
\end{figure*}

\section{Study Design}

Our study aims to explore the following research questions:

\noindent \textbf{RQ1:} Does visual thematic analysis as provided by the proposed methodology and associated tools aid paper selection?

\noindent \textbf{RQ2:} Does visual thematic analysis as provided by the proposed methodology and associated tools aid generation of cross-paper reading strategies?

\noindent \textbf{RQ3:} What are the advantages and disadvantages of the overall proposed visual thematic approach?

To that end, we will ensure our participants are experienced with browsing scientific literature and establishing a reading strategy. Given this experience, and to reduce fatigue in the course of the study, we will not ask the participants to complete a keyword search tasks to contrast for performance and usability against theme-based tasks. However, we will ensure that participants are reminded of this approach using pre-study questionnaires, and we gather insight using stimulated recall semi-structured interviews. 

We designed two user tasks that we ask our participants to carry out to explore and gather insights. Firstly, \textbf{A)}, browse and select a set of research papers using an objective, theme-based overview of a large paper collection. As stated in \textbf{RQ1}, we are interested in analyzing whether taking a theme-based approach, using thematic analysis, aids the selection of papers. This task will also create the basis for investigating \textbf{RQ2}.

Secondly, \textbf{B)}, generate a cross-paper reading strategy using a thematic comparison of a selected paper set. We are interested in facilitating the generation of a reading strategy that considers a set of papers rather than individual strategies for each paper (\textbf{RQ2}).

In addition to their responses of these two user tasks, we analyze perceptions of a theme-based discovery of literature. Throughout the user tasks, we are interested in observing behavior from our participants interacting with our theme-based approach to evaluate whether it allows for high-level insights into research papers, highlighting its advantages and disadvantages as per \textbf{RQ3}.

\subsection{Thematic methodology}

Based on these task requirements, we developed a thematic methodology consisting of two associated thematic tools for the presentation of a large paper collection, and the comparison of a paper set to facilitate the generation of a cross-paper, holistic reading strategy. Our methodology can be summarized in six stages (Figures \ref{fig:f2} and \ref{fig:f3}):

\begin{enumerate}
\item Definition of a large paper collection (Figure \ref{fig:f2}a);
\item Thematic analysis of a large paper collection resulting in a visual thematic map (Figure \ref{fig:f2}b);
\item Upon selection of an individual theme from the thematic map, the top relevant papers are displayed, including the theme location in their content (Figure \ref{fig:f2}c);
\item Six papers are selected by the user on the basis of their interests in investigating these papers further (Figure \ref{fig:f3}a);
\item Papers are represented as theme wheels showing the sequences of themes from start to end, allowing for a cross-paper thematic analysis (Figure \ref{fig:f3}b);
\item An all-inclusive reading strategy based on all six papers is generated by the user (Figure \ref{fig:f3}c).
\end{enumerate}

\begin{figure*}[tbp]
  \centering
  \mbox{} \hfill
  \includegraphics[width=.99\linewidth]{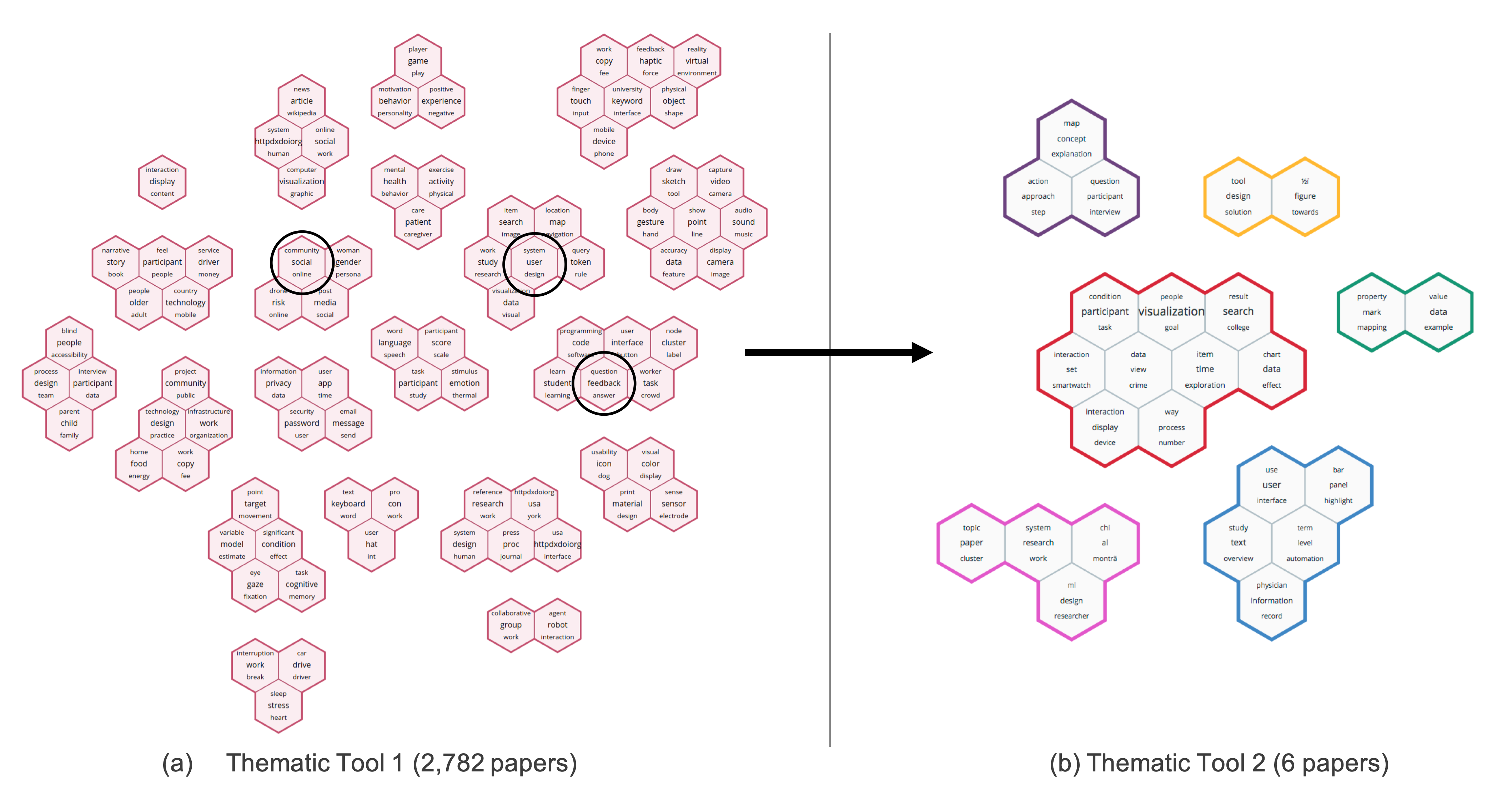}
  \hfill \mbox{}
  \caption{\label{fig:f4}%
           Thematic map evolving from Tool 1, featuring an overview of the paper collection to Tool 2 that features themes from the selected paper set. In the example above, Tool 1 includes 85 topics from 2,782 papers that filters down to only 35 topics from 6 papers in Tool 2.  }
\end{figure*}

\subsection{Thematic Tool 1: browsing and selection}

This tool focuses on Task A, i.e. browsing and selecting within a large paper collection, with the aim to \textbf{cover stages 1-3} of our thematic methodology outlined above (Figure \ref{fig:f2}). An overview of the selected large paper collection is shown using a similarity-based thematic map (Figure \ref{fig:f2}b). This thematic map features clusters of hexagons, each representing a group of similar themes found from the paper collection in a concise, structured and efficient setting. Having these themes rendered as clusters of hexagons allows users to gather insights into the individual themes that are present and investigate which other areas are closely linked and may be of interest.

\subsubsection{Interactivity and aesthetics}
Clicking on a single theme will display a word-cloud representation of the theme, and a listing of the top ten relevant papers, with an explanation for the ordering of the papers: each paper displays its relevance percentage to the theme, and its theme wheel (Figure \ref{fig:f2}c). These are donut chart visualizing which parts of the paper were used to represent the estimation of each theme giving users information regarding where and by how much a theme occurs in the text allowing for better insights, for example, establishing whether the theme is a minor feature of the background section, or consistently used throughout the paper. We chose this method of visualization instead of other types (e.g. bar charts) as these are more aesthetically pleasing and to reinforce relevant percentages \cite{ware2019information}, also incorporating images instead of only text can facilitate understanding as explored in Robb's et al work \cite{robb2015moodsource} \cite{robb2015crowdsourced} \cite{robb2017image}. Finally, we purposefully removed the paper title from the list to create an objective environment, probe discussion, and insights, and bring focus to the theme rather than the title, as discussed in our motivation for this work.

\subsubsection{Implementation}
This tool is implemented using data visualization techniques combined with topic modeling algorithms \cite{blei2012probabilistic} \cite{blei2003latent} that use statistical methods to annotate large archives of documents with thematic information, extracting the common themes among the documents \cite{shivam2020} \cite{blei2012probabilistic}. We used Blei et al. Mallet implementation \cite{mccallum2002mallet} with commonly recommended parameters \cite{boyd2014care} to compile the themes for our study.

We split the individual papers from the collection into equal test chunks. We then use LDA \cite{blei2003latent} applying Gibbs Sampling \cite{mccallum2002mallet}, to uncover the themes and their distribution in the text chunks. We finally compiled the theme distributions for each paper. We visualize the set of uncovered themes in a similarity-based thematic map, using an agglomerative layout process, as described by Le Bras et al. \cite{le2018improving}. This technique allows to visually cluster themes based on their co-occurrences in the papers (Figure \ref{fig:f2}b). We then use theme wheels to present the distribution of a particular theme (selected by the user) throughout the papers presented in a list of the top ten papers per theme (Figure \ref{fig:f2}c).

\subsection{Thematic Tool 2: generation of a reading strategy}

This thematic tool focuses on task B, i.e., generating a cross-paper reading strategy using the selected paper set, with the aim to \textbf{cover stages 4-6} of our thematic methodology outlined above (Figure \ref{fig:f3}). This tool allows for a theme-based analysis of the selected paper set, where we produce a truncated thematic map containing only the themes that are relevant to the papers in the selected set \cite{le2018improving} (Figure \ref{fig:f4}). 

The size of this excerpt map will vary based on the selected paper set. In addition, each paper is represented alongside by its theme wheel representing the structure of papers by visualizing the sequence of themes from start to end (Figure \ref{fig:f3}b).

\subsubsection{Interactivity and aesthetics}
Upon interacting with either of these layouts (the thematic map or the theme wheels), users are presented with a word-cloud to get a detailed description of the themes, emphasizing the relationship between the elements on the screen \cite{yi2007toward}. This allows users to analyze and compare a set of research papers, permitting an in-depth exploration of the consistency and changes of the themes that the paper authors discuss.

The aesthetics of the excerpt map and theme wheels for Tool 2 were designed to emphasize the different theme contributions, distinct themselves from task A (Tool 1), and to make it visually appealing to users.

\subsubsection{Implementation}
Given the selected paper set by participants, the themes covered by each of the selected papers are noted, and this information is extracted from our thematic map from Tool 1, meaning that the number of extracted themes will fluctuate depending on the papers. This creates a smaller thematic map that contains only the relevant themes for these papers. Each of the themes is then re-evaluated in terms of how similar they are to each other using agglomerative clustering algorithms \cite{ackermann2014analysis} \cite{le2018improving} creating our focused thematic map (Figure \ref{fig:f4}). Each cluster of themes is assigned a different color, allowing for a conceptual link between the clusters and 
the theme wheels.

\subsection{Pre-study pilots}
Two pilot studies \cite{turner2005role} were completed to evaluate both tools individually. We evaluated Tool 1’s usability with three participants. This evaluation consisted of a set of tasks followed by the completion of SUS \cite{grier2013system}. The set of tasks comprised of using the tool to explore literature about how users interact with data visualizations, select up to six papers that were believed to be useful in gathering this knowledge and explain reasons for this selection.  Tool 1 received an average usability score of 76 across participants, indicating good interface usability. It also helped us identify usability issues which we corrected.

We focused on Tool 2’s evaluation on the usability of theme wheels. In particular, we looked at how the donut charts were used to investigate literature sources (lecture notes were used due to accessibility). Five participants were given the task of summarizing a set of lectures, which was repeated twice with the order randomized – once using a theme wheel of the whole course and once using a hard copy of the lecture outline materials. This was followed by informal semi-structured interviews in order to gain insights into how participants felt using the two different resources to complete their tasks. We found that the theme-wheels introduced a pictorial representation of the course, allowing for participants to navigate the lecture materials without opening every document and skim-reading each one individually. It, therefore, supported our premise that theme wheels allow for a broad, intuitive, and objective overview of literature sources.

\section{Procedure}
In this section, we detail the steps involved in running our study, including how we recruited participants and coded semi-structured interviews.

\subsection{Data processing}
For our study, we followed the thematic methodology that has been outlined, making use of our two thematic tools. Our large paper collection is made up of five years’ worth of CHI papers, excluding any extended papers, totaling 2,782 papers. 

Papers were then each split into 30 equal text chunks (83,460 in total) and run through LDA \cite{blei2003latent} (as noted in the implementation of Tool 1) and generated 85 themes. This number was settled after exploration sessions and manual adjustments to get detailed themes whilst keeping this number manageable for participants. Figure 5 shows some examples of uncovered themes.

For the second phase, we require the use of Tool 2 which, as described previously, extracts relevant themes based on the selected paper set made by a participant. The size of these excerpt thematic maps varied across participants (n: 10, avg: 28.5, std dev: 7.8, min: 10, max: 37).
\begin{figure}[htb]
  \centering
  \mbox{} \hfill
  \includegraphics[width=.99\linewidth]{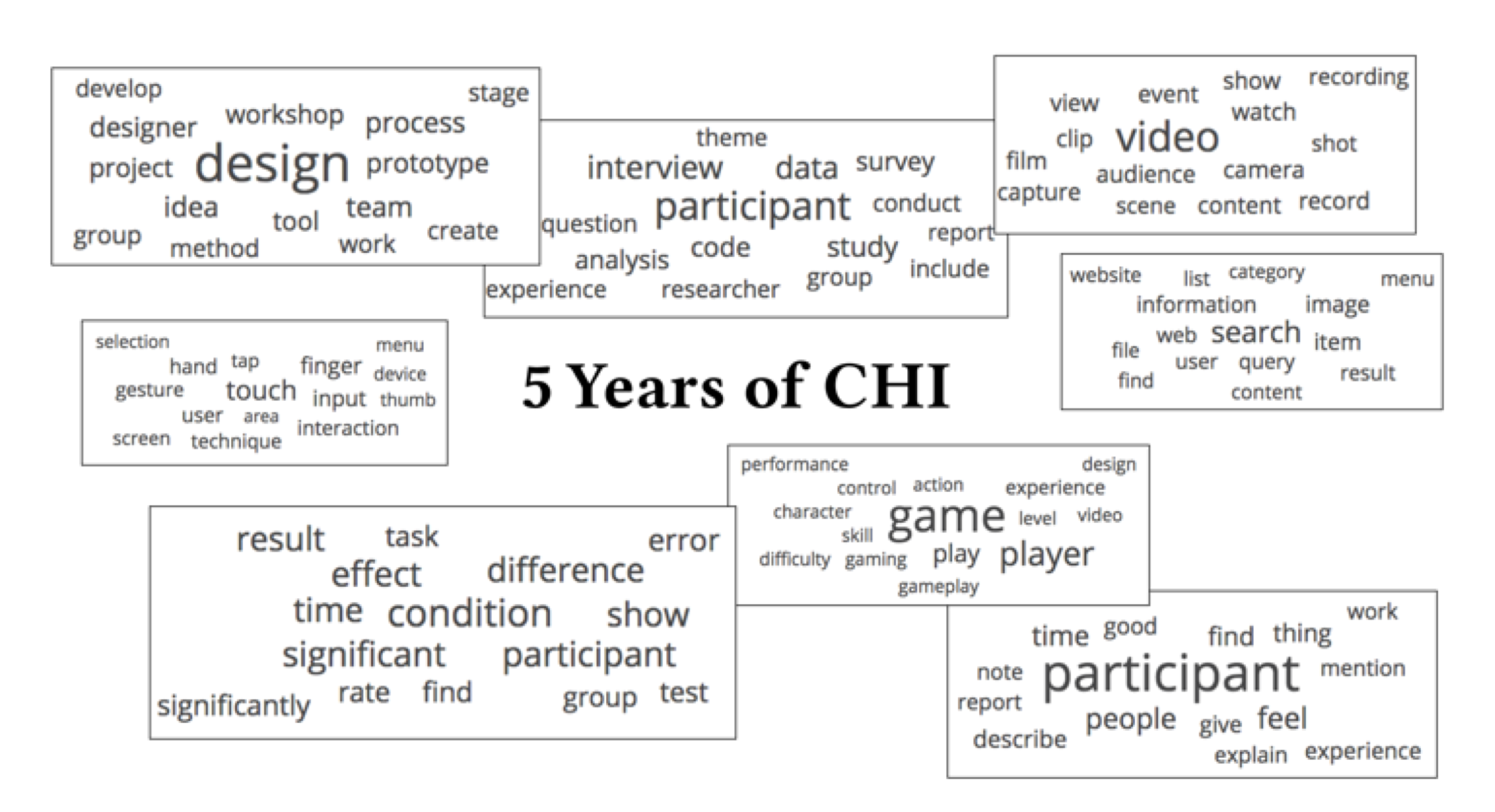}
  \hfill \mbox{}
  \caption{\label{fig:f5}%
           We run thematic analysis on papers from 5 years’ worth of CHI (2,782 papers in total) to give an overview of the research community. A subset of uncovered themes can be seen above.}
           \vspace{-0.8cm}
\end{figure}
\subsection{Participants}
We recruited 10 experienced participants (P1-P10) in total (5 males; 5 females; aged 18-44) using advertisements throughout our organization, which attracted participants across several departments to take part \cite{cairns2008research} \cite{patton1990qualitative}. None reported to be color blind and there was no confusion distinguishing between the colors and shapes used in the tool interfaces. We noticed saturation in the coding of our results as reported in later sections, validating our sample size of participants. 

Using a pre-study questionnaire, we verified that all participants are experienced in using digital repositories to browse for literature sources. These experiences ranged from using digital repositories several times per week (7 participants), at least once a week (1 participant) to less than every 1-2 months (2 participants). The stimulated recall \cite{adams2008questionnaires} of experience was also used during the semi-structured interviews to allow participants to contrast between their experience and our theme-based approach. 

Our study received ethical approval from our institution, and consent was collected from the participants. Every participant was compensated with a \$12 voucher for their time. All the results from this study were anonymized and unlinked.

\subsection{Study}
We divided our study into three stages with two user tasks, aiming to keep participants motivated by breaking down the study into smaller, manageable tasks \cite{cairns2008research}. These stages follow the tasks we describe above, consisting of: A) browse and select 6 papers using Tool 1, B) generate a reading strategy using Tool 2, and C) report on the perception of theme-based literature discovery during a semi-structured interview.

This was accompanied by a scenario within which we asked our participants to place themselves in \cite{jenkins2010putting}, in order to bring focus and context to their tasks and the interview \cite{barter1999use} \cite{jenkins2010putting}. The scenario is as follows: \emph{“You are currently planning an experiment where you will be looking at how people use different websites and what they like and dislike about them. You are interested in using focus groups or interviewing techniques to gather additional insights from your participants. However, you are not sure whether this is the best option for you, so, you want to explore what approaches other similar studies have taken, including how to report on the data gathered.”} This scenario was chosen as it fit into the community of papers that are being displayed and is simple enough that participants are not required to have a background in computing to complete the tasks, allowing for us to reach a more diverse audience \cite{cairns2008research}.

Stage 1 (paper selection using Tool 1) consisted of the first user task, (A), that was performed by participants in their own time 1-3 days prior to the rest of the study. This allowed for the task to feel more relaxing and realistic \cite{jenkins2010putting} and gave the investigators enough time to process data before Stage 2. Participants were also given worksheets to complete, in which they communicated their choice and reasonings. 

\begin{figure}[htb]
  \centering
  \mbox{} \hfill
  \includegraphics[width=.99\linewidth]{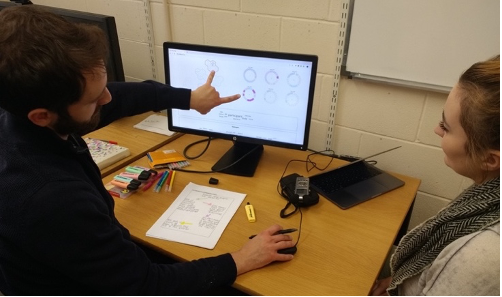}
  \hfill \mbox{}
  \caption{\label{fig:f6}%
           Representation of part 2 of the study. The setup includes the interactive online tool, hard copies of worksheets, markers, highlighters and an audio recorder.}
           \vspace{-0.4cm}
\end{figure}

Stage 2 (reading strategy generation using Tool 2) consisted of the second user task, (B), where participants were shown their selected 6 papers rendered as theme wheels and were asked to analyze and interact with the visualization in order to draw out a plan as to how they would go about investigating the papers further. In particular, we sought to understand their reading strategy in terms of what order they would read the papers and whether they would read only certain parts within the paper. We then revealed the paper titles to the participants and asked them to describe their impression of the title, compared to their analysis of the theme wheel. (Figure \ref{fig:f6} demonstrates the setup).

Finally, Stage 3 (semi-structured interview) sought the participants' opinions and insights about Tool 1 and Tool 2. These interviews lasted no longer than 30 minutes. In particular, we emphasized the interviews towards the participants' usage of the tools, their views on the theme-based approach, their usual procedure with digital repositories, and the contrasts between the two approaches.

The interviews were recorded, with the participants' agreement, and transcripts were then produced for coding.

\subsection{Analysis and coding}

Coding was done by the investigator using computer-assisted qualitative data analysis software \cite{silverman2013doing}. An open coding or inductive approach was used to develop the codebook \cite{corbin2014basics} \cite{fagan2010usability}. After selecting a random transcript, an initial codebook was drawn, and then verified and adjusted on a second transcript. The rest of the transcripts were coded accordingly. A second pass through the data was made to ensure consistency. We found saturation, validating our sample size of experience participants for the study. In addition, we are making the transcribed interviews and analyzed data open for future research in this and other areas (\url{strategicfutures.org/publications} (CC-BY)).   

\begin{figure}[htb]
  \centering
  \mbox{} \hfill
  \includegraphics[width=.99\linewidth]{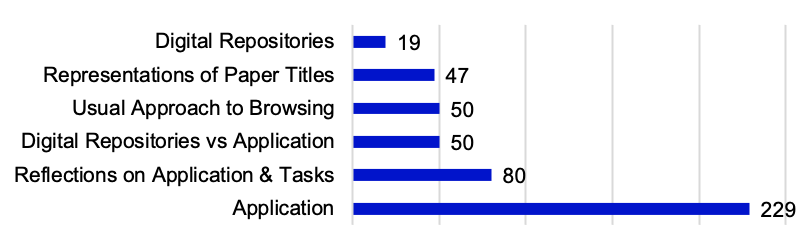}
  \hfill \mbox{}
  \caption{\label{fig:f7}%
          The high-level codes from our semi-structured interviews, measured by analyzing the number of coding references made.}
\end{figure}

\noindent
Our codebook comprises of 6 high-level codes (Figure \ref{fig:f7}):
\begin{itemize}
\item \textbf{Application:} This was the largest topic appearing from our interviews, where participants were asked to think about what they liked and disliked about the tools, how they used the thematic maps and theme wheels, information they thought was missing and how much they interacted with the tools. This brought out any usability issues in terms of features that they did not understand or use; 
\item \textbf{Digital Repositories:} Participants were asked to think about prior experiences using digital repositories allowing them to gather their thoughts as to what they like and dislike about these systems, including what information they thought would be useful to help them navigate and select appropriate texts; 
\item \textbf{Digital Repositories vs Application:} While the previous two codes reflected on our thematic tools or digital repositories separately, this category focuses on the participants’ contrasts between the two approaches; 
\item \textbf{Representation of Paper Titles:} Participants were asked to talk the investigator through each of the paper titles and discuss their reactions to them whilst comparing the title to the themes shown from the theme wheels on Tool 2; 
\item \textbf{Usual Approach to Browsing:} Discussions around digital libraries brought up how participants usually browse literature, giving insights into what they consider useful information about papers. The main criteria used to select papers was also discussed; 
\item \textbf{Reflections on Application \& Task:} Participants were asked to reflect upon how they felt completing the tasks, allowing them to consider whether they would change how they approached the tasks given the knowledge that they now have about the tools.
\end{itemize}


\section{Results and Discussion}
The research questions that were posed in the introduction will now be addressed with design recommendations being made for designers to enhance their literature discovery systems like digital repositories.

\subsection{Selection of papers}
In this section, we discuss RQ1 that focuses on discovering whether a thematic analysis using the proposed theme-based methodology aids paper selection. Analyzing participants worksheets detailing reasons for their paper set, we found that all participants used features of Tool 1 to reason their paper selection. P7 focused on the theme content presented in the thematic map. In addition to this feature, P1, P3, P4, and P10 relied on the calculated theme percentage. P5 combined the theme representation with the theme location within papers. Finally, P2, P6, P8 and P9 made use of all of these features.

During interviews, participants were asked to discuss how they used the thematic tools to complete the tasks and contrast this approach to how they would have usually completed similar tasks with digital repositories. Upon reflection of selecting papers using a thematic approach, P8 pointed out that \emph{“at first it takes a little getting used to because it’s a very different way of considering papers, but it does make you focus on the keywords”}. P7 continues on this point by explaining, \emph{“it’s a little more dynamic, your eyes can first go to keywords of relevance, so it removes that metric of where it is in a list of papers”}. P2 describes that \emph{“it might make narrowing down a scope to a few papers from one hundred and, everybody wants to read as few papers as possible”}.

All ten participants mentioned benefits of this system and it was noted that a thematic approach was \emph{“better than scrolling through a list of titles”} (P3) and helped \emph{“pick out the main themes a lot better than you would get with a list of titles”} (P5). This prompted participants to begin to reflect upon their usual approach, and how much reliance they place on paper titles to help with the selection process as P9 describes \emph{“whenever I’m looking at papers, I probably put a lot of emphasis into the title than I’ve realized”} and P8 recognizes that \emph{“having titles taken away definitely made you think differently and focus a bit more on keywords of what you’re going to get out of it”}. This highlights problems with titles as participants described them as \emph{“always trying to be catchy, they’re attention-seeking and they don’t necessarily say everything”} (P8) resulting in sometimes selecting papers and feeling like \emph{“this isn’t quite what I expected”} (P2).

\noindent
\textbf{Design Recommendation (R1):} We found that all participants appreciated the theme-based analysis and the thematic mapping of the paper collection. We found that similarity-based layouts aided fast theme selection. We would recommend (R1) that designers make use of thematic analysis, consider implementing it alongside their normal search methods, and use a visual similarity-based map, allowing users to easily select themes and explore relevant papers.

Also as mentioned, participants began to reflect upon their usual approach that involved \emph{“quickly scroll through and see different titles”} (P10) or as P3 mentions \emph{“check the titles which will usually get me to discard a few”}. However, when using a thematic approach, participants noted being able to gauge the volume and location of themes as P8 highlights the usefulness of having papers \emph{“ordered using this percentage”} and not placing emphasis on a title because \emph{“a title can be misleading”}. P2 discusses the value of knowing the locations of themes as the visualization \emph{“tells me where this keyword is in the paper… is it in the introduction, which may not be very relevant to me, I might be looking at methods, so this is very useful!”} P5 also mentions this point as \emph{“you can see the location of different topics, you don’t get that in any digital libraries that I know of really or certainly graphically, so yeah, I like that”}. 

When contrasting the thematic approach to participants’ common searching approach, two participants weren’t sure if they would have selected papers in their paper set based on titles alone – “I’m not really sure whether any of these would leap out at me as something that I thought that I would need to read for the kind of research I would like to do” (P5). This point is also mentioned by P7: “I have no idea if that would affect the picking of it if I knew that was the title, it might”.

\noindent
\textbf{Design Recommendation (R2):} Users found that the thematic paper ranking and particularly the graphical, single-theme representation of both the volume and location of a theme within a paper useful. This aided assessment of the relevance and use of the theme within a paper, facilitating the decision to include or not include this paper within the paper set. We recommend that designers provide these meta-data (paper ordering and theme volume), which are often generated by search engines \cite{barter1999use} \cite{beel2009google} but not normally made available to users, as they aid selection.

As we have seen, a thematic approach facilitates the selection of research papers, but it also allows for a more objective method to filtering papers that resulted in participants selecting papers that they believe would not have been selected if a traditional approach was being used. This is due to the functionality allowing papers to be filtered by themes, and the ability to show the volume and location of the theme.

\subsection{Development of a reading strategy}
In this section, we focus on RQ2, that poses whether the thematic methodology and associated tools can aid the generation of a cross-paper reading strategy given a selected paper set. In the second phase of the study, participants were asked to consider the selected paper set and develop a reading strategy. From worksheets that participants described their strategies on, we found that six participants ordered papers for reading based on how much they contained the main themes that the participant was most interested in. 

P2 describes their answer as \emph{“looking at the color coding and looking at the general themes in the papers”} whereas P3 notes they would \emph{“investigate the purple bits because there are a lot them, there are almost three whole purple donuts!”} Others, such as P5, described their approach to investigating the paper set as \emph{“scrolling along here [the theme wheels] and then seeing which ones [themes from thematic map] light up and how that relates to the papers that I picked”} to find out \emph{“what the predominant color is”}.

Participants also used the theme-based overviews to eliminate papers that after having a closer look at, no longer seemed as relevant as noted by P5 – \emph{“Paper 1, I didn’t end up using because I thought it was more specifically for musical learning and it was quite good I realized that, so it wasn’t used”}. P7 summarizes by stating, \emph{“I think the visual aspect is helpful because you can almost kind of quickly quantify what a single paper is about whereas with Google Scholar it’s kind of just a list of links”}.

Planning out a reading strategy allows for participants to focus on what they want to get out of each paper to solve a problem. This was highlighted by many participants when they discuss their usual approach to the discovery of literature as being \emph{“very disorganized”} (P5) or \emph{“surfing from paper to paper”} (P8). This highlights the piecemeal approach that is often adopted using common searching techniques, as digital repositories do not allow for a paper set to be considered and evaluated, only individual sources.

\noindent
\textbf{Design Recommendation (R3):} We found that when given a side by side comparison of the multi-theme representations of sequences of themes within a paper set, participants could formulate a cross-paper reading strategy, ordering paper sections that they plan to read based on the quantity and positions of themes within each individual paper, promoting a coherent approach to investigating the sources. We recommend that designers facilitate the comparison of a paper set using visualizations of each papers’ sequences of themes.

It is clear from discussions with participants, that a combination of the thematic map and the theme wheels were used to develop a reading strategy. Due to the clustering in the thematic maps, similar themes were grouped together. Participants mentioned this functionality as it \emph{“provides a link to something that might be worth exploring”} (P2) but three out of the ten participants also commented on having difficulties to \emph{“ﬁnd the exact keywords that I noted [in the previous tasks]”} (P6).

In order to reduce this problem from occurring, we can imagine a closer integration between Tool 1 and Tool 2. This could be done using visual explanations \cite{le2018improving} to animate the evolution of the thematic map from Tool 1 to Tool 2, allowing for users to trace \cite{gregor1999explanations} interesting themes and see how the tools pull out relevant information.

\noindent
\textbf{Design Recommendation (R4):} Our result show that participants felt they would have benefited from a closer link between the thematic map of the paper collection provided in Tool 1 and the more focused thematic map provided for the selected paper set in Tool 2 (see Figure \ref{fig:f4}). We recommend that the thematic maps of the paper collection and the user’s paper selection are tightly integrated (e.g. the provision of common highlighting, multiple selections or interactive transitions).

Based on the evidence presented from the in-depth interviews, we have found that not only does following a thematic approach aid the generation of a reading strategy, but often a strategy that takes into consideration a set of papers as a whole rather than traditional approaches where users adopt a more piecemeal strategy.
\subsection{Levels of insight into papers}
In this section, we discuss RQ3 that focuses on the advantages and disadvantages of the overall thematic approach that has been proposed. Exploring RQ1 and RQ2, we have seen that a thematic approach to discovery and analysis of literature gives insights into the structure, author keywords and sequence of themes as mentioned by participants whilst discussing advantages and disadvantages of following a thematic approach. 

P2 describes being able to \emph{“pick out bits of a paper that were on a particular topic that I might want to focus on, so I could see, oh that’s a bit of wafﬂe, so I can skip through that”} while P7 mentions that \emph{“I like how you can see the progress through a paper like that, being able to see how the topics change or don’t change”}. Theme wheels allowed participants to easily identify paper sections (e.g. introduction, background or conclusion), enabling them to map their knowledge and experiences with research papers. 

With such a focus on themes, six out of ten participants found that they interpreted themes differently to the content, which was brought to light when the titles were uncovered. For example, P7 describes this as \emph{“I just saw privacy and thought data privacy and I don’t know if this is actually what this is on or if it’s more actual physical privacy? But I was thinking more data protection online, so yeah, I was surprised by that”}. Le Bras et al. work recommends interactivity incorporated into the map for increased user confidence and engagement as participants can then interrogate the process and understand the information at their own pace \cite{le2018improving}. Therefore, giving users the ability to change the level of detail being displayed in our thematic maps could lead to a deeper understanding of not only individual themes but also clusters of themes.

\noindent
\textbf{Design Recommendations (R5):} We found that some participants would have liked to have been able to obtain a deeper understanding of particular themes at both the paper selection and reading strategy generation stages. We recommend that designers explore hierarchical thematic analysis techniques \cite{Griffiths2003hierarchical} to allow users with different levels of knowledge to investigate themes at multiple levels of abstraction.

Nine out of our ten participants noted being surprised by at least one paper title when they saw the titles at the end of the tasks. This is emphasized when participants were asked to explain their reactions where P1 mentions that \emph{“the first two, no, I would never ever imagine it was that”} and P3 mentions that a paper was \emph{“meant to be for interviewing techniques since the tags were interview, data, survey but the title is nothing like that”}. P2 found that the titles were \emph{“totally different but still useful”} Whilst P9 reflects on their approach by mentioning \emph{“I used the keywords quite a lot, so the title was quite different, so it was quite surprising”}. 

Uncovering the titles of the papers right at the end of the study highlights our previous point that titles are only one to two lines long so cannot be expected to reflect the full content. Therefore, by introducing thematic overviews of the content, participants could see the progression of themes from start to finish, giving insights into the tools and techniques used but sometimes lacking in giving context to the research. For example, eight out of our ten participants selected a paper titled, Investigating the Suitability of the Asynchronous, Remote, Community-based Method for Pregnant and New Mothers \cite{prabhakar2017investigating}. This title came as a surprise to all eight of these participants, like P5 who said it \emph{“surprised me a bit. I didn’t see anything in here [the application] that made me think of that”} or P7 who said, \emph{“I definitely had no idea that this was what the paper would be about”}.

Our chosen algorithm aims to uncover the most common themes in a whole corpus of text. It is, therefore, not surprising that pregnant and new mothers do not come out as a major theme in HCI community. This did not cause issues to participants for their task, as that paper discusses qualitative methods such as focus groups and interviews, meeting the given scenario and task. If participants had been given the task of understanding the context of papers, it would then be likely that they would have struggled to grasp this information from the theme wheels alone. 

During the interviews, participants were asked whether they thought a thematic approach could be a replacement of current digital repository systems or if it would be more valuable as an add-on feature. Only two participants thought that our thematic methodology could replace current systems, with the other eight participants believing that this approach would be best as an add-on feature. P9 reasoned this as \emph{“getting used to new systems is quite difﬁcult, so it would be good to have that alongside”} or as P8 suggests, \emph{“people are so stuck in their ways, so I don’t know how open-minded people would be”}.

Participants began to describe how they would use current systems with a layer of thematic information added. P2 mentions that \emph{“I would probably start with this [interface] to get me to a place where I think I am ready to look at the text and start looking at the abstracts then and progress from there”} while P1 states, \emph{“I really love this interface, it's perfect for the ﬁrst screening but then you need something else [such as access to digital repositories]”}.

\noindent
\textbf{Design Recommendation (R6):} Participants appreciated the integrated thematic approach and its visual representation and interface. However, during the study, the participants clearly expected the title and abstract fields to be also available and would appreciate a combination of approaches. We recommend that designers incorporate visual thematic analysis tools with traditional title-abstract search methods to allow users to seamlessly switch between and combine approaches to get both theme and context information.

As we have seen, based on results from our semi-structured interviews, there are advantages and disadvantages to the overall proposed thematic approach. Advantages included the ability to have a visual representation of a large collection of papers, see the sequences of themes from start to finish in a paper and visually compare a paper set in order to aid the generation of a cross-paper reading strategy. The main disadvantages highlighted by participants were not having an integrated environment with traditional information such as titles and abstracts available to them, but they appreciated that this process did allow for them to reflect upon their common approach to the discovery of literature and question their reliance on commonly used information for their reading strategies.


\section{Conclusions}

In this paper, we present a study exploring the effects of a new visual methodology and complementary toolset that helps users browse, select and develop holistic reading strategies. We principally focus on whether our proposed approach enriches paper selections, facilitates the development of coherent reading strategies, and allows them to develop high-level holistic reading strategies. To explore these aspects, we carried out a three-phase qualitative study using scenario-based, semi-structured interviews that were designed to probe insight into to the use of our methodology and tools. We investigated participants' approaches, user behaviors, and reactions using our thematic methodology and contrasted them to their experiences with common digital repositories.

We believe that our results indicate that adopting a visual thematic methodology encourages a more objective approach to browsing and selecting papers. Participants chose papers that they thought they would definitely otherwise would have not selected and, following selection of paper sets, participants used a combination of visual thematic maps and theme wheels to develop theme-based, cross-paper reading strategies. In addition, participants found that the multi-theme paper visualizations gave useful insights into the structure, ordering, frequency and commonality of themes, allowing participants to quickly gain an overview of content, authors’ writing styles and focus.

We make six recommendations aimed at assisting designers that wish to enhance or develop visual thematic tools and methodologies that will help users quickly and efficiently explore digital repositories. We certainly believe that such tools should be closely integrated with existing approaches to provide complementary, rather than replacement functionality, in order to encourage a more holistic and objective approach to developing reading strategies.

Finally, we hope the insights, visualizations, methodology, tools and recommendations proposed in this paper will encourage discussion in the community and catalyze the development of new visual thematic-based approaches to developing interfaces to a wide variety of digital repositories, including for example storing video, audio, and multimedia data for educational, entertainment and governmental applications.

\subsubsection*{Acknowledgments}
The authors would like to thank the participants for their time and insightful discussions. The data generated for this study can be accessed on request, please email the authors for further details. Finally, visualisations of corpora, open algorithms and data (CC-BY), similar complementary tools \cite{methven2014research} \cite{padilla2012digital} \cite{methven2015don}, and related work can be access at \url{strategicfutures.org}.

\bibliographystyle{eg-alpha-doi}
\bibliography{references}


\end{document}